\documentclass[10pt,showpacs,showkeys,preprint]{revtex4-1}
\usepackage{bm}
\usepackage{amssymb}
\usepackage{amsmath}
\usepackage{graphicx}
\usepackage{multirow}
\usepackage{MnSymbol}
\usepackage{wasysym}
\usepackage{stmaryrd}
\usepackage[outline]{contour}
\usepackage[dvipsnames]{xcolor}
\usepackage{float}
\usepackage[mathlines]{lineno}

\definecolor{olive}{RGB}{182,187,37}

\begin{document}

\title{Atomic ionization by multicharged ions interpreted in terms of poles
in the velocity complex space}
\author{J. E. Miraglia}
\date{\today }
\date{}

\begin{abstract}
We study the single ionization of hydrogen and helium by the impact of a
highly-charged Coulomb projectile. To interpretate the cross section we
introduce a diagonal Pad\'{e} approximant. We find that the use of Pad\'{e}%
[4,4] describes very well the Continnum Distorted Wave Eikonal Initial State
theory within its range of validity. The nodes of the denominator of the Pad%
\'{e} approximant give rise to four poles in the velocity complex plane: two
in the upper plane and their conjugate in the lower plane. The dependence of
these poles with the projectile charge can be reasonably fitted to give a
closed-form for the ionization cross section, resulting a scaling very near
to the one of Janev and Presnyakov. The experiments available were described
very well in its entire velocity range with the use of a Pad\'{e}[8,8],
having four poles in the upper plane and their conjugate in the lower plane.
We conclude that the poles of the Pad\'{e} approximant seem to have all the
information of the total ionization cross section
\end{abstract}

\keywords{ionization, atoms, scaling, multicharged ions}
\pacs{34.50Gb}
\maketitle

\affiliation{Instituto de Astronom\'{\i}a y F\'{\i}sica del Espacio. Consejo Nacional de
Investigaciones Cient\'{\i}ficas y T\'{e}cnicas} 
\affiliation{Departamento de F\'{\i}sica. Facultad de Ciencias Exactas y Naturales.
Universidad de Buenos Aires. \\
Casilla de Correo 67, Sucursal 28, {C1428EGA} Buenos Aires, Argentina.}


\section{Introduction}

With the advancement of the big accelerators, the subject of ionization of \
atoms by highly-charged ions has developed notably. It was accompanied with
the development of several quantum and classical theories, for example, the
distorted waves \ such as the Continuum Distorted Wave Eikonal Initial State
(CDW for short) \cite{Fainstein1988}, the Classical Trajectory MonteCarlo
(CTMC) \cite{Olson1977}, close coupling calculations such as the
Basis-Generator Methods (BGM) \cite{Ludde1996},\ amongst others. For
many-electron atoms  the situation is certainly\ very complex due to several
mechanisms involved, including several electron transitions. But even with
simple targets such as hydrogen and helium, the experiments are limited.
Still there is not a theory that can predict with certainty\ the ionization
of these simple systems for a large charge of the projectile $Z$ and\ for a
given impact velocity $v$. In fact, the only theoretical method that has
accompanied consistently most of the experiments, has been the CTMC; its
simplicity, ubiquity and applicability makes this classical method a
favorite tool to deal with large $Z.$ The quantum CDW theory is a very
useful and reliable tool, but its validity is reduced to the
intermediate-high energy region. On the other side, the BGM \ gives account
of the experiments almost in the whole velocity range. Its calculation
involves a high degree of computing, and its extension to high $Z$ seems to
be quite complicated. High projectile charges and small-velocity impacts are
still an unsolved problem.

Impulsed by the hadron therapy to deal with cancer the theoreticians are
forced to deal with ionization of ADN molecules by charged projectile such
as C$^{6+}$. Therefore, the challenge of dealing with large $Z$ \ has been
resurfaced,\ but now within the more complex field of molecular targets. To
deal with this challenge, some models reduce the problem to a sum of
ionization cross sections of the atoms composing the molecules \cite%
{Mendez2020}. More refined approaches take also into account the geometry of
the molecule, but still the problem of dealing with high $Z$ persists.
Recently\ Kirchner and collaborators \cite{Ludde2020} have designed a method
to extrapolate the BGM from $Z=4$ to tackle the problem of ionization of
uracil by C$^{6+}$ because the BGM complicates as $Z$ increases$.$

Some scalings have been already designed to deal with this problem proposing
a normalization of the velocity $v$ and cross section $\sigma (Z,v)\ $with $Z
$ trying to unveil a universal curve which permits to extrapolate to high $Z$%
. Based on the first Born \ approximation, a family of scalings is possible,
by writing%
\begin{equation}
\frac{\sigma (Z,v)}{Z^{\alpha }}\propto \frac{Z^{2-\alpha }}{v^{2}}=\left( 
\frac{Z^{1-\alpha /2}}{v}\right) ^{2}=\frac{1}{\xi ^{2}},  \label{10}
\end{equation}%
where we can identify different options, for example%
\begin{equation}
\left\{ 
\begin{array}{lll}
\alpha =2, & \xi _{B}=v, & \text{Born,} \\ 
\alpha =1, & \xi _{JP}=\frac{v}{\sqrt{Z}}, & \text{Janev and Presnyakov, Ref.%
\cite{Janev1980},} \\ 
\alpha =4/3,\ \ \  & \xi _{M}=\frac{v}{Z^{1/3}},\ \ \ \  & \text{Ref.\cite%
{Montenegro2013},} \\ 
\alpha =1.2,\  & \xi _{Mol}=\frac{v}{Z^{1/3}}, & \text{for molecules Ref.%
\cite{Mendez2020}.}%
\end{array}%
\right.   \label{12}
\end{equation}%
The range of validity of a given theoretical method is generally expressed
in terms of the corresponding $\xi .$ It is well known that the Born scaling
holds when the Sommerfeld parameter is small i.e. $\xi _{B}=v>>Z.$ The most
\ popular scaling at intermediate impact energies is the one of Janev and
Presnyakov $\xi _{JP}$ \cite{Janev1980} . It $\ $was originally introduced
to deal with dipole transitions but \ today it has been extended to a large
variety of inelastic direct processes with \ great success. We will prove
that the CDW strongly relates to this scaling. There are also other
scalings: $\xi _{M}$ proposed by Montenegro \textit{et al }\cite%
{Montenegro2013} which work better at lower velocities, Gillespie has also
devised an scaled exponential universal factor with the argument \ $\xi
_{JP}^{2}$ \cite{Gillespie1982}.$\ $A complete study of the different
expressions and approaches was published by Kaganovich \textit{et al} \cite%
{Kaganovich2016}.

In this article we examine the ionization of simple atoms, hydrogen and
helium, by impact of high charges. By high charges we mean $Z$ as large as $%
30$ for hydrogen and $56$ for helium. Our strategy is novel and creative. We
propose that the ionization cross section can be expressed in terms of a
particular Pad\'{e} approximation which is essentially a coefficient of
polynomials in term of velocity $v$. The zeros of the one in the
denominator\ correspond to the poles\ of the cross section in the complex
plane of the velocity.\ We find that these poles in the CDW theory move with 
$Z\ \ $following a certain pattern to the point that we can find an
approximated closed-form \ for large $Z$, which we find it is related to the
Janev and Presnyakov parameter $\xi _{JP}$. We reduce the problem to just
two moving poles in the upper complex plane and their conjugate in the lower
one, determining the cross section in terms of the $v$ and $Z$. \ At the
very end, the great challenge of this strategy is the description of the
experimental cross sections. We find that for protons, antiprotons and $%
He^{++}$ impact the experimental data \ can be very well replicated \ with
four poles in the whole energy range. Inner shell ionization can also be
described with four poles as well as O$^{8+}$ on helium. In conclusion, this
article proposes that the ionization cross section can be reduced to the
knowledge of four poles in the velocity complex plane.

The idea that the ionization cross section could have poles in the velocity
complex plane should not be peculiar: Green operators have poles in the
imaginary component in the k-space, and resonances are explained in terms of
poles in the energy complex plane. In similar fashion the maximum of the
ionization cross section is here read as the presence of a pole near to the
real axis and a threshold of ionization as the \ competing modulus of all
the poles.

The work is organized as follows. After introducing the Pad\'{e} approximant
in section 2, we proceed to find the poles of the Born and CDW theoretical
methods and finally we localize the position of the poles\ projected by the
experimental data. Atomic units are used.

\section{THEORY}

\subsection{Experimental and numerical data set}

We should first define our universe of work which is the ultimate target of
our study. In Table 1 we resume an experimental data set (EDS) of ionization
cross section $\sigma ^{\exp }(Z,v)$ from different laboratories totalizing
125 (203) experimental values for hydrogen (helium)\ for different charges $Z
$\ and impact velocities $v$, including antiproton impact ($Z=-1$). Details\
of the references as well as charges and velocities considered are displayed
in the Table. There are some other experiments by impact of projectile that
cannot be considered punctual as required by the theory. This Table should
not be considered complete, we simply resume the most relevant ones obtained
in a more or less systematic way. To have an idea of the measurement
spectrum, in Fig 1a (2a) we show each experimental value in a $Z-v$ plot for
hydrogen (helium). Helium for obvious reasons presents\ a more complete
panorama.

The only theory that we use in this article is the CDW. In a similar fashion
we build a numerical data set (NDS) of 245 (198) values of hydrogen (helium)
target by the impact of different $Z$ and $v$.\ In Figure 1c (2c), we show
each theoretical calculation in the $Z-v$ plot for hydrogen (helium). Some\
values can be found in the literature:\ for hydrogen in \cite{Miraglia2019}
and for helium in \cite{Miraglia2020} for charges $Z=-1$ to $8.$ We also
extended the calculation for negative charges up to $Z=-8$ ($Z=-4$) for
hydrogen (helium). Of course, except antiprotons, negative charges are
unrealistic, but let us understand the process in a wider range.

Before proceeding we point out that for helium target, we did a full
calculation of its continuum state expanding in spherical harmonics of the
potential obtained with the depurated inversion model\ \cite{Mendez2016}.
This potential reads%
\begin{eqnarray}
V_{He}(r) &=&-\frac{1.0764}{r}\exp (-2.79681r)(1+0.62529r)  \label{20} \\
&&+\frac{0.0764}{r}\exp (-18.3544r)(1+9.8617r)-\frac{1}{r},  \notag
\end{eqnarray}%
warranting at least four significant figures of the binding energy, and
three figures for the mean values of  $\left\langle 1/r\right\rangle ,$ $%
\left\langle r\right\rangle \ $and $\left\langle r^{2}\right\rangle $ as
compared with Hartree Fock. 
\begin{table}[tbph]
\caption{Experimental data set (NDS). Projectile energies are in MeV/amu and
N is the number of points.}
\label{table2}%
\begin{tabular}{|ccccc|ccccc|}
\hline\hline
Target & Z & \ \ Ref.\ \  & \ Energy range \  & \ \ N \ \  & Target & Z & \
\ Ref.\ \  & \ Energy range \  & \ \ N \ \  \\ \hline
H & -1 & \text{\cite{Knudsen1995}} & 31-800 & 19 & He & 3 & \text{\cite%
{Knudsen1984}} & 640-2310 & 3 \\ 
H & 1 & \text{\cite{Rudd1985}} & 30-1500 & 11 & He & 4 & \text{\cite%
{Knudsen1984}} & 190-2310 & 6 \\ 
H & 1 & \text{\cite{Shah1987}} & 9-75 & 12 & He & 5 & \text{\cite%
{Knudsen1984}} & 190-2440 & 4 \\ 
H & 1 & \text{\cite{Shah1981c}} & 38-1500 & 27 & He & 6 & \text{\cite%
{Knudsen1984}} & 640-2260 & 3 \\ 
H & 2 & \text{\cite{Shah1981c}} & 31-550 & 17 & He & 7 & \text{\cite%
{Knudsen1984}} & 1440-2260 & 2 \\ 
H & 2 & \text{\cite{Shah1988}} & 19-64 & 12 & He & 8 & \text{\cite%
{Knudsen1984}} & 640-2260 & 4 \\ 
H & 3 & \text{\cite{Shah1982}} & 57-387 & 14 & He & 24-54 & \text{\cite%
{Berg1992}} & 3600 & 8 \\ 
H$_2$/2 & 11 $\rightarrow$ 22 & \text{\cite{Berkner1978}} & 1100 & 8 & He & 
6-44 & \text{\cite{McGuire1987}} & 1400 & 7 \\ 
H$_2$/2 & -1 & \text{\cite{Hvelplund1994}} & 13-29 & 5 & He & 26-44 & \text{%
\cite{Datz1990}} & 1000 & 10 \\ 
He & -1 & \text{\cite{Andersen1990}} & 40-2890 & 26 & He & 22-37 & \text{%
\cite{Datz1990}} & 500 & 10 \\ 
He & -1 & \text{\cite{Hvelplund1994}} & 13-500 & 23 & He & 9-31 & \text{\cite%
{Datz1990}} & 250 & 23 \\ 
He & 1 & \text{\cite{Shah1985}} & 64-2380 & 17 & He & 5-16 & \text{\cite%
{Datz1990}} & 100 & 8 \\ 
He & 1 & \text{\cite{Rudd1985}} & 15-5000 & 16 & He & 8 & \text{\cite{Wu1995}%
} & 27-72 & 1 \\ 
He & 2 & \text{\cite{Shah1985}} & 50-1585 & 16 & He & 6 & \text{\cite%
{Schlachter1981}} & 310-1140 & 20 \\ 
He & 2 & \text{\cite{Knudsen1984}} & 640-2310 & 3 & He & 10-14 & \text{\cite%
{Tonuma1985}} & 1050 & 5 \\ 
He & 3 & \text{\cite{Shah1985}} & 50-390 & 11 & He & 6, 8 & \text{\cite%
{Wu1996}} & 70-250 & 15 \\ \hline\hline
\end{tabular}%
\end{table}
\begin{figure*}[tbh]
\centering
\includegraphics[width=0.95\textwidth]{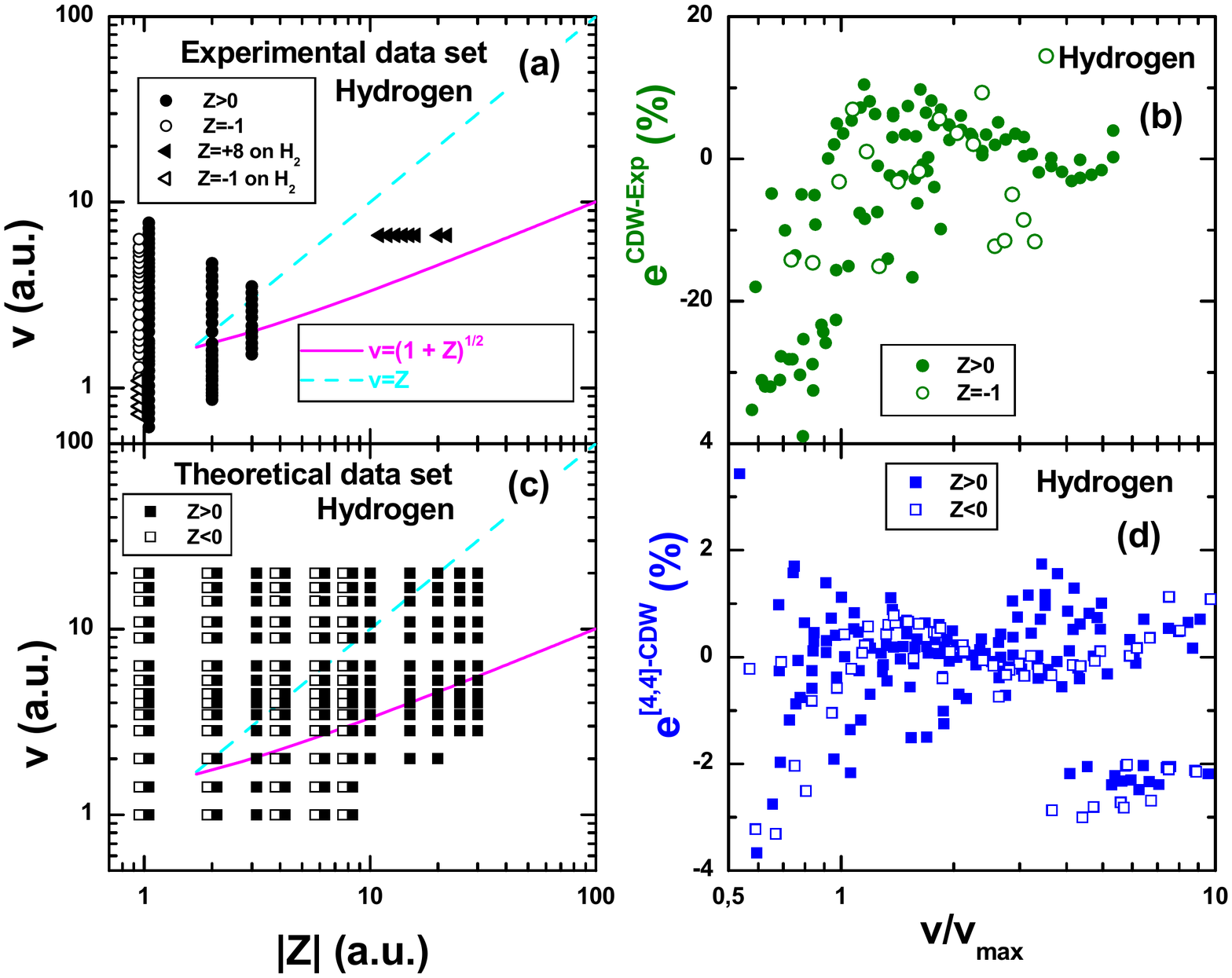}
\caption{(Color online) Hydrogen target. Figure (a) experimental data set
and (c) numerical data set in a Z-v diagram. Figure (b) relative error of
the CDW versus the experiments as defined in Eq.(5) in terms of $v/v_{max}$.
Figure(d) relative error of $\protect\sigma ^{\lbrack 4,4]}$ versus the CDW
calculation as defined in Eq.(18) in terms of $v/v_{max}$.}
\end{figure*}
\begin{figure*}[t]
\centering
\includegraphics[width=0.95\textwidth]{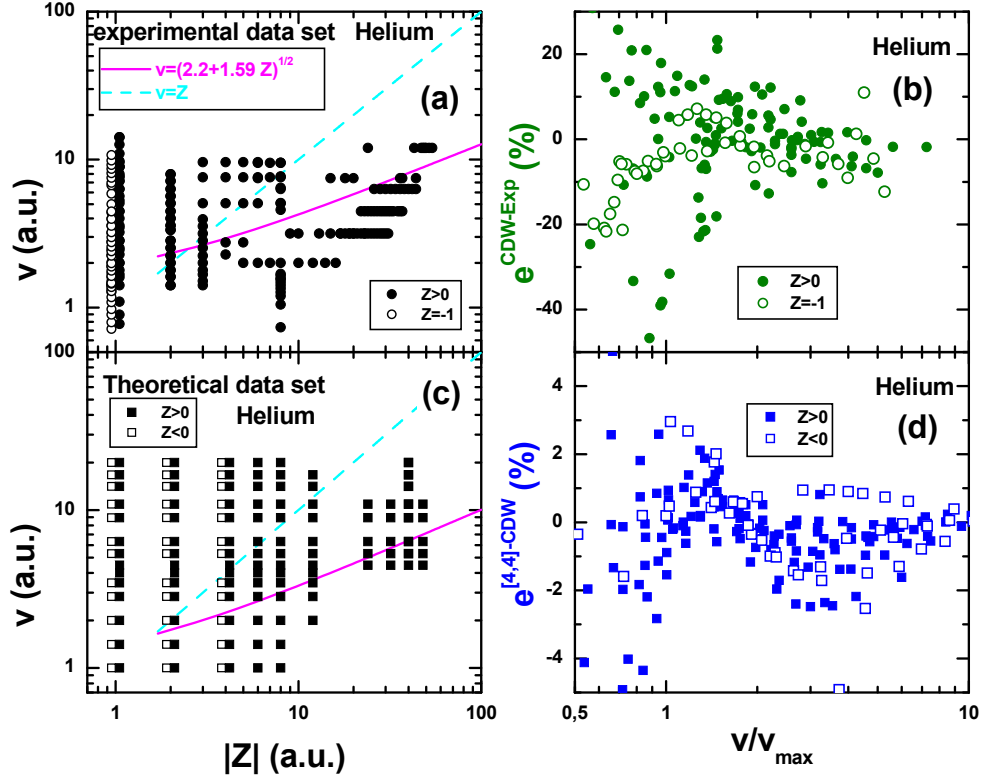}
\caption{(Color online) Helium target. Figure (a) experimental data set and
(c) numerical data set in a Z-v diagram. Figure (b) relative error of the
CDW versus the experiments as defined in Eq.(5) in terms of $v/v_{max}$.
Figure (d) relative error of $\protect\sigma ^{\lbrack 4,4]}$ versus the CDW
calculation as defined in Eq.(18) in terms of $v/v_{max}$. }
\end{figure*}
\subsection{Definition of the validity regime}

From our NDS we can easily obtain the velocity $v_{\max }\ $where the cross
section is maximum,$i.e$. 
\begin{equation}
\sigma _{\max }(Z)=\sigma (Z,v_{\max }(Z)).  \label{345}
\end{equation}%
\ \ In \ Fig 3a (3b) we show the values of $\ v_{\max }^{2}(Z)$ and $\sigma
_{\max }(Z)\ $for hydrogen (helium). Notably for $Z>0,$ $\sigma _{\max }(Z)$
and $v_{\max }^{2}(Z)\ $behaves linearly with $Z,\ $and the dependence can
be fitted \ approximately as,

\begin{eqnarray}
v_{\max }^{2} &\approx &\left( v_{\max }^{B\ \ }\right)
^{2}+c_{1}Z\rightarrow c_{1}Z,\ \ \text{and}\ \   \label{350} \\
\sigma _{\max } &\approx &\frac{Z^{2}\sigma _{\max }^{B\ \ }}{1-c_{2}\sqrt{Z}%
+Z\frac{\sigma _{\max }^{B\ \ }}{c_{3}}}\rightarrow c_{3}Z,  \label{355}
\end{eqnarray}%
where \ $v_{\max }^{B\ \ }$ =$1.\ \ (1.25)$ is the \ velocity where the Born
approximation $\sigma ^{B\ \ }$ is maximum for hydrogen (helium), and $%
Z^{2}\sigma _{\max }^{B\ \ }$ is the value of Born cross section at the
maximum with $\sigma _{\max }^{B}=\sigma ^{B}(v_{\max }^{B\ \ })=7.63\ (3.67)
$ for hydrogen (helium). The rest of the coefficients are found to be:\ \ $%
c_{1}=1\ (1.59),\ c_{2}=0.48\ (0.54),$ and $c_{3}=7.32\ (5.42)\ \ $\ for
hydrogen (helium). Relation (\ref{350}) is fundamental, since it lets us
introduce a criterion to define the validity of the CDW-theory, and that is
when%
\begin{equation}
v\gtrsim v_{\max }\text{,\ \ or\ \ }\xi _{JP}\gtrsim \sqrt{c_{1}+\frac{1}{Z}%
\left( v_{\max }^{B\ \ }\right) ^{2}}\rightarrow \sqrt{c_{1}}\   \label{370}
\end{equation}%
which is stated in terms of the parameter Janev and Presnyakov $\xi _{JP}\ $%
\ and not to the Sommerfeld criterion defined as $\xi _{B}>>Z.\ $It is
important to note that negative charges have also a linear dependence with $Z
$ but with a different slope. The magnitude $v_{\max }$ as defined in Eq.(%
\ref{350}) is displayed in red solid-line in Figure 1a, for hydrogen and 1b
for helium to indicate the intermediate energy region in contrast with $\xi
_{B}=v=Z\ $\ in blue dashed-line\ which sets the lower border of validity of
the Born approximation.

Now we can quantify the validity of the CDW by contrasting this theoretical
prediction with the experiments in terms of $v/v_{\max }.$ In Fig 1b (2b) we
display the relative error defined as%
\begin{equation}
e^{CDW-\exp }(\%)=\frac{\sigma ^{CDW}(Z,v)-\sigma ^{\exp }(Z,v)}{\sigma
^{\exp }(Z,v)}\times 100,  \label{375}
\end{equation}%
for all the universe of experiments listed in Table 1 (or situated in Fig 1a
for hydrogen and 1b for helium). where we can observe that the range of
applicability of the CDW is indeed restrained to $v\gtrsim v_{\max }$. From
the figures, we can state that the CDW predicts the experiments with $\pm
20\%$ at $v\sim v_{\max }$, converging for larger $v$. At the same time the
error explodes for $v<v_{\max }.$ Therefore $v_{\max }$ , as defined in Eq.(%
\ref{350}),\ rests as a firm standpoint of the intermediate energy.

\subsection{The Pad\'{e} approximant}

First we divert the atomic ionization cross section $\sigma (Z,v)\ $with the
help of the diagonal Pad\'{e} approximant $P_{m,m}$ and the asymptotic
limits $L(v)\ $by writing%
\begin{eqnarray}
\sigma _{Z}^{[m,m]}(Z,v) &=&Z^{2}L(v)P_{m,m}\left( Z,v\right) ,  \label{100}
\\
P_{m,m}\left( Z,v\right)  &=&\frac{\overset{m}{\underset{\mu =0}{\dsum }}%
n_{\mu }(Z)v^{\mu }}{\overset{m}{\underset{\mu =0}{\dsum }}d_{\mu }(Z)v^{\mu
}},  \label{110} \\
L(v) &=&\frac{A}{v^{2}}\ \log (1+Bv^{2}),  \label{131}
\end{eqnarray}%
being $Z^{2}L(Z,v)$ the correct asymptotic limit, and therefore it is
required that%
\begin{equation}
\underset{v\rightarrow \infty }{\lim }\ P_{m,m}\left( Z,v\right) =1.
\label{141}
\end{equation}%
Written in this way $L(Z,v)$ is finite at the origin:\ $\ L\left( v\right)
\rightarrow AB,$ as $v\rightarrow 0.\ $After a series of trials we have
considered it convenient to use the following contracted form for the
diagonal $P_{2n,2n}$ approximant%
\begin{equation}
P_{2n,2n}\left( Z,v\right) =\underset{j=1}{\overset{n}{\dprod }}\frac{v^{2}}{%
(v-v_{+j})(v-v_{-j})},  \label{165}
\end{equation}%
where $v_{\pm j}~$ are the position of the poles in the velocity complex
plane defined as%
\begin{equation}
v_{\pm j}=v_{jr}\pm iv_{ji}\ .  \label{180}
\end{equation}%
We cast on the real and imaginary components of the poles, $v_{jr}$ and $%
v_{ji},$ the dependence on the projectile charge $Z$; that is we expect: $%
v_{\pm j\ r/i}=v_{\pm j\ r/i}(Z).$\ All the information then is being
reduced to the position of $n$ poles in the complex upper plane of the
projectile velocity, \ $v_{ji}>0,\ $and their conjugate in the lower plane,\ 
$v_{ji}<0$. The imaginary part warrants that there is no divergence at real
velocities, and this is the reason why\ we have chosen the particular
expression given by (\ref{165}). By construction the term $P_{2n,2n}$%
satisfies the condition (\ref{141}) and at the threshold it behaves as $%
v^{2n},i.e.$%
\begin{equation}
\sigma ^{\lbrack 2n,2n]}(Z,v)\underset{v\rightarrow 0}{\rightarrow }\
Z^{2}AB\ v^{2n}\underset{j=1}{\overset{n}{\dprod }}\frac{1}{\left\vert
v_{j}\right\vert ^{2}}.  \label{210}
\end{equation}%
It rests now to find the best value of $n$ determining the degree of the Pad%
\'{e} approximant to be used. The problem is that there is not a solid
knowledge about the actual behaviour of the ionization cross section in this
region. In the literature we find different extrapolated expressions,
contradictory to each other,  other showing a certain level of uncertainty.
Just to illustrate the spread: the recommended values of Rudd \cite{Rudd1985}%
\ behave as $v^{2D}$ with $D=0.907\ (1.52)$ for hydrogen ( helium) but the
successful expression of Gillespie decays exponentially \cite{Gillespie1982}.

One illuminating study is the theory of inner-shell ionization developed in
the seventieths by Basbas \textit{et al } \cite{Basbas1973} which has been
largely used with great effectiveness. This theory is based on the simple
first Born that the authors found to have a behaviour $v^{8}$ in the region
where $v<<v_{\max },$ which is the region that we precisely need to access.
Inspired in this, we propose $\ n=4$ leading to a $P_{88}$ that is%
\begin{equation}
\sigma ^{\lbrack 8,8]}(Z,v)=Z^{2}L(v)P_{8,8}\left( Z,v\right) ,  \label{212}
\end{equation}%
which can be seen as a simple product of two $P_{44,}$ or a product of \
four $P_{22}.$ In this way we expect our expression to be high- and
low-energy properly bond.

The first part of this article will concentrate in finding the poles of the
CDW theory, which we expect to be valid for $v\gtrsim v_{\max }$. In that
case we\ resume the calculation to the use of just $P_{44}$ having \ the
correct high energy bond, and a $v^{4}-$behavior at the origin. Which is
supposed to be\ incorrect, but it is worthless to include more poles to
refine the expression in a region were the CDW\ fails. We will find that $%
\sigma ^{\lbrack 4,4]}(Z,v)$ is enough, and reserved $\sigma ^{\lbrack
8,8]}(Z,v)$ to investigate the experimental data which is a much more
demanding task. We can also read $\sigma ^{\lbrack 4,4]}(Z,v)$ as a
particular case of $\sigma ^{\lbrack 8,8]}(Z,v),$ namely%
\begin{equation}
\sigma ^{\lbrack 4,4]}(Z,v|v_{1},v_{2})=\sigma ^{\lbrack
8,8]}(Z,v|v_{1},v_{2},v_{3}=0,v_{4}=0).  \label{214}
\end{equation}%
Thus, we can then identify the two extra poles commanding the intermediate
and the threshold. By threshold we mean the region starting from few keV
where the projectile can be considered a heavy particle describing a
straight line trajectory.

\subsection{The Born Poles}

The first test to check that our expression is appropriate, is to examine
the first Born approximation which is supposed to be correct as $%
Z\rightarrow 0.$ Further, the Born approximation lets us determine the
values of $A$ and $B$ \ of the asymptotic limit expression $L(Z,v)$. Thus,
we obtain $A$ $=3.52\ (6.11)$, and $B=62.1\ (3.22)\ $ for hydrogen (helium).
The values of $A$ agree with the ones used by \cite{Rudd1985} and \cite%
{Voitkiv1988},\cite{Voitkiv1998}. While dealing with $\sigma ^{\lbrack
4,4]}(0,v)$ we have observed that $v_{1r}$ was positive while $v_{2r}\ $\
was always negative and $v_{2r}\sim -v_{1r}$. so we decided to set 
\begin{equation}
v_{2r}=-v_{1r}\ ,\   \label{305}
\end{equation}%
and reduce to three the number of free parameters to be fitted. We will come
back to this point The values of\ $v_{1r},~v_{1i},$ and $v_{2i}$ are
displayed in table 2 in the row corresponding to $Z=0$ for hydrogen and
helium\ targets, as indicated. This Table also displays the poles
corresponding to $\sigma ^{\lbrack 8,8]}(0,v)\ $ using the same data set and
imposing also $v_{4r}=-v_{3r}.$ At intermediate energies the results are
very similar, but at lower velocities they differ: $\sigma ^{\lbrack
4,4]}\propto v^{4},$ \ while $\sigma ^{\lbrack 8,8]}$ $\propto v^{8}.$

\subsection{The CDW-EIS poles}

Next, we proceed to find the values of the poles $v_{1}$ and $v_{2}$
governing the CDW theory. For that, end we use the NDS\ shown in Figs 1a and
2a. As in Born approximation, we considered $v_{2r}=-v_{1r}:\ $the agreement
with the numerical data is not altered substantially if we let them vary
freely. One interesting feature of the Pad\'{e} approximant so-defined is
that the sum of the residues of all the poles of the boundary-corrected
function $Q_{44}\left( v\right) =P_{4,4}\left( v\right) -1,$ is%
\begin{equation}
S_{4,4=}\dsum\nolimits_{j=1}^{2}\left( \text{Residue[}\left.
Q_{44}\right\vert _{v_{j}}\text{]+Residue[}\left. Q_{44}\right\vert
_{v_{j}^{\ast }}\text{]}\right) \text{=}2(v_{1r}+v_{2r}),  \label{306}
\end{equation}%
so by setting condition (\ref{305})\ \ then $S_{4,4}=0.$ As a consequence of
this, the integration on a closed curve $C=|v|\rightarrow \infty ,$ produces
also a null value%
\begin{equation}
I_{4,4}=\doint\nolimits_{|v|\rightarrow \infty }Q_{44}(v)\ dv=0.  \label{307}
\end{equation}%
Another way to visualize Eq.(\ref{307}) is to expand $Q_{44}$ for large
values of the complex magnitude $v$ to give%
\begin{equation}
Q_{44}(v)\underset{v\rightarrow \infty }{\rightarrow }\ \frac{%
2(v_{1r}+v_{2r})}{v}+\mathcal{O}\left( \frac{1}{v^{2}}\right) ,  \label{309}
\end{equation}%
by setting $v_{1r}+v_{2r}=0,$ we obtain null circulation given by Eq.(\ref%
{307}). If we require: $\sigma ^{\lbrack 2n,2n]}(Z,v)\rightarrow Z^{2}L(v)$
as the \textit{real }magnitude\ $v\rightarrow \infty ,$ we are restraining
the limit to just the \textit{real} axes, but the condition (\ref{305})
generalizes this limit in the entire velocity \textit{complex} velocity
magnitude $|v|\rightarrow \infty .$

\begin{figure*}[t]
\centering
\includegraphics[width=0.95\textwidth]{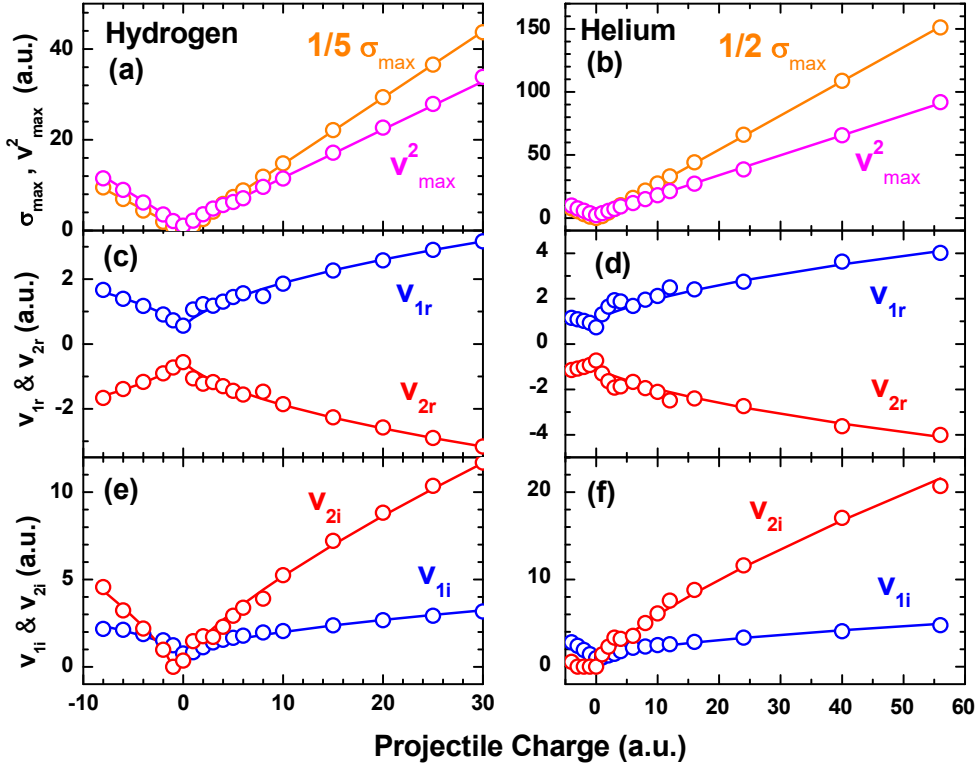}
\caption{(Color online) Fig (a) maximum value of the ionization cross
section $\protect\sigma _{max}$, and $v_{max}^{2}$ for hydrogen target as
defined in Section II.B. Fig (b) similar variables for Helium. Figures (c)
and (e), real and imaginary parts of the poles: $v_{1r}$ and $v_{2r}$, and $%
v_{1i}$ and $v_{2i}$, respectively, for hydrogen target, as a function of
the projectile charge. Figures (d) and (f) similar variables for helium
target. }
\end{figure*}

In Figure 3c ( 3d), we plot the values of$\ v_{1r}$ and $v_{2r}\ $for
hydrogen (helium), and in Figure 3e (3f ) the values of $v_{1i}$ and $v_{2i}$
for hydrogen ( helium) as a function of the impinging charge $Z$. We have
observed that the prediction of the Pad\'{e} \ $P_{4,4}\ $approximant is
excellent in our range of interest here. To illustrate that our Pad\'{e}
approximant $\sigma ^{\lbrack 4,4]}$ gives a quite precise description of
the CDW\ results, we plot in Fig 1d the relative errors with respect to the
full numerical results\ $\sigma ^{CDW}$ defined 
\begin{equation}
e^{[4,4]-CDW}=\frac{\sigma ^{\lbrack 4,4]}(Z,v)-\sigma ^{CDW}(Z,v)}{\sigma
^{CDW}(Z,v)}\times 100  \label{310}
\end{equation}%
as a function of $v/v_{\max }\ $for all the cases of the NDS. For $v/v_{\max
}\gtrsim 1,$ errors are less than 4\%, but most of the them are around or
even less than 2\% which is of the order of the numerical uncertainties of
the $\sigma ^{CDW}$ calculation. Similarly, in Fig 2d we show the equivalent
relative error for helium. \ The agreement of $\sigma ^{\lbrack 4,4]}$ with
the numerical results cover almost all the velocity range of our interest,
except at the threshold $v<<v_{\max \ }$where the $v^{4}-$dependence imposed
by our $P_{4,4}\ $is no longer correct.

There are a lot of interesting rules in the position of the poles which
could lead to some physical interpretation. The first observation -which
perhaps is the main finding of this work- is that the components of the
poles $v_{1r},\ v_{1i},$ and $v_{2i}$\ obey certain patterns to the point
that we can find a reasonably fitted closed-form, say $V_{1r},$ $V_{1i}$ and 
$V_{2i}$,\ expressed as%
\begin{equation}
\left\{ 
\begin{array}{l}
V_{1r}(Z)=-V_{2r}(Z)\approx k_{1}\sqrt{k_{2}+Z} \\ 
V_{1i}(Z)\approx k_{3}\sqrt{k_{4}+Z} \\ 
V_{2i}(Z)\approx k_{5}\sqrt{\left( k_{6}+Z\right) ^{3/2}}%
\end{array}%
\right. \text{,}  \label{330}
\end{equation}%
where \ $k_{1}=0.560\ (0.526),\ \ k_{2}=1,(6.24)\ \ k_{3}=0.560\ (0.636),\
k_{4}=2.20\ (2.80),\ \ k_{5}=0.921\ (1.05),\ $\ and $k_{6}=0.278,(0.0),$ for
hydrogen (helium)\ target.$\ \ $The fact that Im[$v_{\pm 2}(0)$]=0\ does not
present any problems since the divergence occurs at negative (unphysical)
impact velocities.\ For negative charges, \ $-8<Z<0$, \ we can also fit
better the NDS\ and the pole description is very good and the errors are
small,\ and this is probably because capture is absent, the process of
ionization becomes simpler and the poles are enough to describe the
mechanism in a cleaner way.

The Poles can now be visualized at the approximate positions%
\begin{equation}
V_{\pm j}(Z)=V_{jr}\pm iV_{ji},\ \ j=1,2\ ,  \label{340}
\end{equation}%
and in this way, we can obtain an approximate ionization cross section $%
\sigma _{V}^{[4.4]}(Z,v)\ $defined through Eq.(\ref{165}) with the \ poles
at $V_{1}$ and $V_{2}$ instead.\ What is interesting is that$\ \sigma
_{V}^{[4.4]}(Z,v)\ $\ has a closed-form. For large values of $Z\ (i.e.\
Z>\max (k_{2}.k_{4},k_{6}))$ it tends to%
\begin{equation}
\sigma _{V}^{[4.4]}(Z,v)\rightarrow Z^{2}L(v)\frac{v^{2}}{\left\vert v-k_{1}%
\sqrt{Z}+ik_{3}\sqrt{Z}\right\vert ^{2}}\frac{v^{2}}{\left\vert v+k_{1}\sqrt{%
Z}+ik_{5}\sqrt{Z^{3/2}}\right\vert ^{2}}.  \label{380}
\end{equation}%
Note the important role of the imaginary part of the second pole behaving as 
$Z^{3/4},$ playing a decisive role for large $Z,$ provided of course that $%
v>v_{\max }.$ Another interesting aspect of Eq.(\ref{380}) is that we can
introduce $\xi _{JP}$ and normalize $\sigma _{V}^{(V_{1},V_{2})}$ to $Z,$ to
have%
\begin{equation}
\frac{\sigma _{V}^{[4.4]}(Z,v)}{Z}\rightarrow \frac{A\log (1+Bv^{2})\ \xi
_{JP}^{2}}{\left\vert \xi _{JP}-k_{1}+ik_{3}\right\vert ^{2}\left\vert \xi
_{JP}+k_{1}+ik_{5}\ \ Z^{1/4}]\right\vert ^{2}}.  \label{390}
\end{equation}%
And this is almost the Janev and Presnyakov scaling. The \ factor $\xi
_{JP}\ $comes up in natural form\ as a consequence of a particular movements
of the poles in the complex plane. Its role is relevant at the maximum, $i.e.
$ around\ $\xi _{JP}=k_{1},$ that is precisely the region where this scaling
was originally intended \ to work at by Janev Presnyakov\ in its original
paper \cite{Janev1980}. But the role of $\xi _{JP}$\ is not fully decisive
because the imaginary part of the second pole $\ k_{5}Z^{1/4}$\ breaks a
perfect scaling with $\xi _{JP}.$

\begin{table}[tbph]
\caption{ Real $v_{1,r}=-v_{2,r}$, $v_{3,r}=-v_{4,r}$ , and imaginary parts $%
v_{1,i}$, $v_{2,i}$, $v_{3,i}$, and $v_{4,i}$ of the Poles correspoding to
the $P^{[8,8]}$ in atomic units fitting the experimental data shown in
Figure 3}
\label{table3}%
\begin{tabular}{|r|c|ccccccc|}
\hline\hline
$Z$ \  & Target & $v_{1,r}$ & $v_{1,i}$ & $v_{2,i}$ & $v_{3,r}$ & $v _{3,i}$
& $v_{4,i}$ \ \  &  \\ \hline
-1 & H & \ \ 0.5869 \ \  & \ \ 1.213 \ \  & \ \ 0 \ \  & \ \ 0.4573 \ \  & \
\ 0.4545 \ \  & \ \ 0 \ \  &  \\ 
(Born) 0 & H & \ \ 0.560 \ \  & \ \ 0.7645 \ \  & \ \ 0.352 \ \  & \ \ 0 \ \ 
& \ \ 0 \ \  & \ \ 0 \ \  &  \\ 
(Born) 0 & H & \ \ 0.5013 \ \  & \ \ 0.7873 \ \  & \ \ 0 \ \  & \ \ 0.2763 \
\  & \ \ 0.3117 \ \  & \ \ 0 \ \  &  \\ 
1 & H & \ \ 0.1336 \ \  & \ \ 1.304 \ \  & \ \ 0 \ \  & \ \ 0.9053 \ \  & \
\ 0.6008 \ \  & \ \ 0 \ \  &  \\ 
2 & H & \ \ 0.4715 \ \  & \ \ 1.920 \ \  & \ \ 0 \ \  & \ \ 1.147 \ \  & \ \
0.5287 \ \  & \ \ 0 \ \  &  \\ 
-1 & He & \ \ 0.7558 \ \  & \ \ 1.673 \ \  & \ \ 0 \ \  & \ \ 0.6787 \ \  & 
\ \ 0.3120 \ \  & \ \ 0 \ \  &  \\ 
(Born) 0 & He & \ \ 0.5417 \ \  & \ \ 0.8020 \ \  & \ \ 1.387 \ \  & \ \ 0 \
\  & \ \ 0 \ \  & \ \ 0 \ \  &  \\ 
(Born) 0 & He & \ \ 0.7367 \ \  & \ \ 1.024 \ \  & \ \ 0 \ \  & \ \ 0.3021 \
\  & \ \ 0.2214 \ \  & \ \ 0 \ \  &  \\ 
1 & He & \ \ 1.207 \ \  & \ \ 1.273 \ \  & \ \ 0 \ \  & \ \ 0.5744 \ \  & \
\ 0.3449 \ \  & \ \ 0 \ \  &  \\ 
2 & He & \ \ 1.051 \ \  & \ \ 2.053 \ \  & \ \ 0 \ \  & \ \ 1.357 \ \  & \ \
0.8662 \ \  & \ \ 0 \ \  &  \\ \hline\hline
\end{tabular}%
\end{table}
\begin{figure*}[t]
\centering
\includegraphics[width=0.95\textwidth]{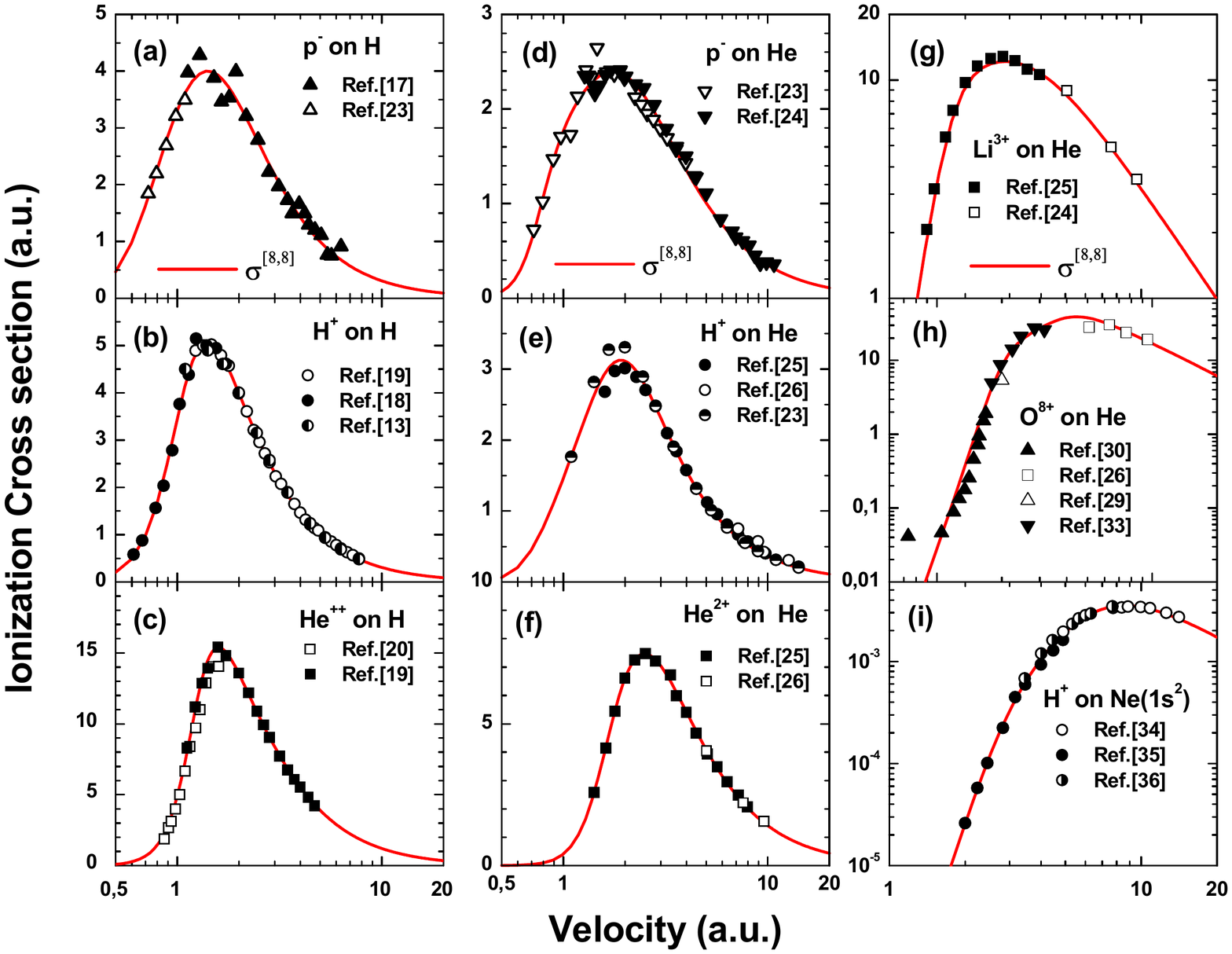}
\caption{(Color online) Ionization cross section of hydrogen, helium and
K-shell of Neon as a function of the impact velocity by impact of diferent
projectiles as indicated. The theory $\sigma ^{[8,8]}$ denoted in read solid lines, 
and symbols are the experiments}
\end{figure*}

\subsection{The experimental Poles}

By observing the excellent performance of the Pad\'{e} approximant, we are
encouraged to extend this scheme to reproduce the experiments to the best
that our model permits. In this case we need to accede to small velocities:
i.e. $v<v_{\max }$ region, that we have discarded when describing the theory
valid for $v\gtrsim v_{\max }$. In that case the $P_{44}$ was enough. But
here we pretend our expression to cover all the experimental range and
therefore we must resort to the original $P_{8,8}$ which we suppose that is
more appropriate.

The great problem is that the experimental data are very limited. Even for
the most popular systems the measurements are very sparce. A direct fitting
of the data is possible for few cases such as $\ H^{+}$\ and $He^{++}$
impact, most of them carried out by Shah,\ Gilbody and Knudsen and
collaborators. \ For antiproton impact on hydrogen the fitting can be
possible by resorting to the Hvelplund \textit{et al }measurements on
molecular hydrogen divided by 2\ \cite{Hvelplund1994} to have some values
for $v<v_{\max }$. In this way we have a minimum number of pivots to use the
minimization algorithm. In any case, the spread of the experimental errors
introduce some noise in the fitting procedure. To guide the algorithm we
have sometimes needed to introduce some theoretical values at very large
impact velocity in the region where there is no experiment at all, but the
theory is expected to hold.

The expression\ $\sigma ^{\lbrack 8,8]}(Z,v)$ depends in principle on eight
variables corresponding to the real and imaginary part of the 4 poles in the
upper plane We follow the same condition of Eq(\ref{305}) and impose 
\begin{equation}
v_{2r}=-v_{1r},\text{\ \ \ \ }v_{4r}=-v_{3r},\   \label{410}
\end{equation}%
reducing the problem to 6 parameters. In similar fashion to Eqs(\ref{306})
and (\ref{307}), the condition (\ref{410}) produces $S_{88}=I_{88}=0.$

\ We were able to fit just three experimental cases: p$^{-},$ H$^{+}$ and He$%
^{++}$ on hydrogen and helium as shown in Figure 4 (a--f). Calculations
based on $P_{8,8}$ gives a very good agreement with the experiments.
Stimulated with this performance we go further facing the system Li$^{3+}$
and O$^{8+}~$\ on Helium where we can put together a reasonable set of
experimental values for the fitting procedure to work. As shown in figure 4g
and 4h, \ the agreement is excellent even for $v<<v_{\max }$ where the
capture channel plays the dominant role. The Pad\'{e} $P_{8,8}$ can also
apply to inner-shell \ as shown in Fig 4i where we fit the \ experiments of
ionization of inner shell of neon. The agreement is again excellent. And we
tested this performance for other inner shell cases as well.

Inspecting the values of components of the poles in Table II\ we find
notable behaviours, for example in some cases\ $v_{2i}=v_{4i}=0$ \ along
with the condition (\ref{410}) constrain the problem to find just 4
parameters. Recall that the fact that the imaginary part is null does not
present any problems since the divergence occusr at negative (unphysical)
impact velocities.


\section{SUMMARY}

We have studied the \ single ionization of a punctual Coulomb charge on
hydrogen and helium. To deal with impact velocities $v\gtrsim v_{\max },$\
we have proposed that the cross section can be separated in an asymptotic
dependence times a Pad\'{e} $P_{44}$ written in terms of our poles: two in
the upper velocity complex plane and their conjugate in the lower plane. By
setting condition (\ref{305}) we could find the components of the poles for
a numerical data set consisting of 443 numerical CDW full calculations for
different values of $Z$. The agreement was estimated less than 4\% for $%
v\gtrsim v_{\max }$. Finally, we deal with the experimental data by using an
appropriate Pad\'{e} $P_{88}$ instead. For the few experiments available
that we could fit, we have found a very good agreement with the data. The
reduction of the physics of the problem to find the poles in the velocity
complex plane is appealing. To find the logic of the dependence with $Z\ $\
of these poles would be a stimulating advance. It would lead to a different
reading of the physical processes involved.

\section{Bibliography.}

{}

\end{document}